\documentclass[aip,jcp]{revtex4-1}

\usepackage{amsfonts}
\usepackage{bm}
\usepackage{setspace}
\usepackage{graphicx}
\usepackage{color}
\usepackage{verbatim}
\usepackage{float}
\usepackage{array}
\usepackage{booktabs}
\usepackage{amsmath}
\usepackage{cleveref}
\usepackage{amssymb}
\usepackage{graphicx}
\usepackage[version=3]{mhchem}
\usepackage[T1]{fontenc}
\usepackage{subfigure}
\usepackage{subfig}
\usepackage{xr}
\usepackage{threeparttablex}
\usepackage{amsmath}

\begin{document}
\title{Charge Transfer Excitation Energies From Ground State Density Functional Theory Calculations}
\author{Yuncai Mei}
\affiliation{Department of Chemistry, Duke University, Durham, North Carolina 27708, USA}
\author{Weitao Yang}
\email{weitao.yang@duke.edu}
\affiliation{Department of Chemistry, Duke University, Durham, North Carolina 27708, USA}
\affiliation{Key Laboratory of Theoretical Chemistry of Environment, School
of Chemistry and Environment, South China Normal University, Guangzhou
510006, China}
\date{\today}

\begin{abstract}
    Calculating charge transfer (CT) excitation energies with high accuracy and low computational
    cost is a challenging task.
    Kohn-Sham density functional theory (KS-DFT), due to its efficiency
    and accuracy, has achieved great success in describing ground state
    problems. To extend to excited state problems, our group
    recently demonstrated an approach with good numerical results to calculate
    low-lying and Rydberg excitation energies of an $N$-electron system
    from a ground state KS or generalized KS 
    calculations of an $(N-1)$-electron system via its orbital energies. 
    In present work, we explore further the same methodology to describe 
    CT excitations. Numerical results from this work show that performance 
    of conventional density functional approximations (DFAs) is not as 
    good for CT excitations as for other excitations, due to the 
    delocalization error. Applying localized orbital scaling correction (LOSC)
    to conventional DFAs, a recently developed method in our group to effectively reduce the delocalization
    error, can improve the results. Overall, the performance of this methodology is better than
    time dependant DFT (TDDFT) with conventional DFAs.
    In addition, it shows that results from LOSC-DFAs in this method reach similar accuracy to other methods,
    such as $\Delta$SCF, $G_0W_0$ with Bethe-Salpeter equations, particle-particle 
    random phase approximation,
    and even high-level wavefunction method like CC2.
    Our analysis show that the correct $1/R$ trend for CT excitation can be captured
    from LOSC-DFA calculations, stressing that the application of DFAs with minimal delocalization error
    is essential within this methodology. This work provides an
    efficient way to calculate CT excitation energies from ground state DFT.
\end{abstract}

\maketitle
\section{Introduction}
Charge transfer (CT) excitation energy calculation has gained much attention in materials and biological sciences
\cite{akimov2013theoretical,bredas2004charge,mason1959charge}. In a 
spatially separated donor-acceptor (D-A) system with intermolecular distance $R$, the CT excitation process involves electron
transition from donor molecule (D) to acceptor molecule (A), and the CT excitation energy
at asymptotically large $R$ is given by
\begin{equation}
    \label{eq:CT}
    E^{\text{CT}} = \text{IP(D)} - \text{EA(A)} - 1/R,
\end{equation}
where IP(D) and EA(A) are the ionization potential (IP) of donor and electron affinity (EA) of acceptor,
and the $1/R$ term comes from the Coulomb attraction.
In practice, calculating CT excitation energy with high accuracy and
low computational cost is interesting and challenging. Among the theoretical developments,
Kohn-Sham density functional theory (KS-DFT) \cite{kohn1965self,kohn1965self,parr1994density}, as a useful
method to ground state problems with great success during the past decades, has not been
well justified for direct application to excited states.
However, many methods developed in the framework of DFT have
been reported. The time-dependant DFT (TDDFT) has been widely
used for single low-lying excitation energy calculation
\cite{runge1984density,laurent2013td,dreuw2005single,marques2006time}.
However, TDDFT faces the challenge in describing CT excitation problems
with commonly used density functional approximations (DFAs)
\cite{dreuw2005single,gritsenko2004asymptotic,maitra2017charge}.
It is known that TDDFT with application of conventional functionals, 
such as local density approximation (LDA),
generalized gradient approximation (GGA) and hybrid functionals,
normally underestimates the CT excitation energies, and
cannot capture the correct asymptotically $1/R$ dependence at a long distance.
Such failure of TDDFT is attributed to two factors.
One is related to the energy bandgap between
the HOMO (highest occupied molecular orbital) and LUMO 
(lowest unoccupied molecular orbital) that
is used to approximate the fundamental gap $(\rm{IP - EA})$,
the leading term in Eq. \ref{eq:CT} at large separation distance.
\cite{autschbachChargeTransferExcitationsTimeDependent2009a,
gritsenkoAsymptoticCorrectionExchange2004a,
maitraUndoingStaticCorrelation2005,
kummelChargeTransferExcitationsChallenge2017,
tozer2003relationship,mooreChargeTransferChargeTransferLikeExcitations2015,
srebroDoesMoleculeSpecificDensity2012}
The other reason is that exchange-correlation kernel, $f_{xc}$, from conventional 
functionals vanishes exponentially at long distance, leading to the absence of the $1/R$ behavior
\cite{autschbachChargeTransferExcitationsTimeDependent2009a,
dreuwLongrangeChargetransferExcited2003a}.

On the physical meaning of HOMO and LUMO energies, it has been proved that for
conventional DFAs, when the approximate 
exchange-correlation energy
$E_{\rm{xc}}[\rho(\mathbf{r})]$ is an 
explicit and continuous functional
of electron density, either local or nonlocal, 
the Kohn-Sham HOMO energy $\varepsilon_{\rm{HOMO}}^{\rm{KS}}$
is equal to the chemical potential for electron
removal (as a prediction of the corresponding DFA);
namely ${\varepsilon_{\rm{HOMO}}^{\rm{KS}}} = 
\left(\frac{\partial E}{\partial N}\right)_V^- = \mu^-$.
Similarly, the Kohn-Sham LUMO energy ${\varepsilon_{\rm{LUMO}}^{\rm{KS}}}$
is equal to the chemical potential for electron addition;
namely ${\varepsilon_{\rm{LUMO}}^{\rm{KS}}} = 
\left(\frac{\partial E}{\partial N}\right)_V^+ = \mu^+$.
\cite{cohenFractionalChargePerspective2008a}
When the approximate exchange-correlation energy
$E_{\rm{xc}}[\rho(\mathbf{r},\mathbf{r'})]$ is 
an explicit and continuous functional of the non-interacting
one-electron density matrix $\rho_s(\mathbf{r},\mathbf{r'})$,
the generalized Kohn-Sham HOMO energy $\varepsilon^{\rm{GKS}}_{\rm{HOMO}}$
is equal to the chemical potential for electron removal;
namely ${\varepsilon_{\rm{HOMO}}^{\rm{GKS}}} = 
\left(\frac{\partial E}{\partial N}\right)_V^- = \mu^-$,
and the generalized Kohn-Sham LUMO energy $\varepsilon^{\rm{GKS}}_{\rm{LUMO}}$
is equal to the chemical potential for electron addition;
namely ${\varepsilon_{\rm{LUMO}}^{\rm{GKS}}} = 
\left(\frac{\partial E}{\partial N}\right)_V^+ = \mu^+$.
\cite{cohenFractionalChargePerspective2008a}
Note that in the case of $E_{\rm{xc}}[\rho_s(\mathbf{r},\mathbf{r'})]$,
the Kohn-Sham (or the optimized effective potential) 
HOMO and LUMO eigenvalues are different from the chemical potentials. 
\cite{cohenFractionalChargePerspective2008a}
When a DFA satisfies the linearity condition for fractional
numbers of electrons,
\cite{perdewDensityFunctionalTheoryFractional1982,
yangDegenerateGroundStates2000}
the chemical potential are related
to the predicted ionization $I$ and electron affinity $A$;
namely $\mu^- = -I$ and $\mu^+ = -A$. For finite and small
systems, the deviation of the DFA from the linearity conditions
is the delocalization error of a DFA. 
\cite{mori-sanchezLocalizationDelocalizationErrors2008b} 
Therefore accurate prediction
of $I$ and $A$ can be obtained from the (G)KS HOMO and LUMO
energy, when the DFA has minimal delocalization error. However,
because commonly used DFAs have significant delocalization error,
the (G)KS HOMO underestimates $I$ and the (G)KS LUMO overestimates
$A$, leading to the underestimation of energy bandgaps
\cite{mori-sanchezLocalizationDelocalizationErrors2008b} 
and also the CT excitation energies
\cite{autschbachChargeTransferExcitationsTimeDependent2009a,
gritsenkoAsymptoticCorrectionExchange2004a,
maitraUndoingStaticCorrelation2005,
kummelChargeTransferExcitationsChallenge2017,
tozer2003relationship,mooreChargeTransferChargeTransferLikeExcitations2015,
srebroDoesMoleculeSpecificDensity2012}.

One attempt to cure the problem is to reduce the delocalization
error by introducing hybrid functionals with a fraction of
Hartree-Fock (HF) exchange, which is known to produce
localization error\cite{cohen2011challenges,cohenInsightsCurrentLimitations2008a}. In addition,
it has been discussed that the full HF exchange can recover the $1/R$ trend in principle at long distance\cite{dreuw2005single}.
With this aspect, applying range-separated functional with system-dependent tuned parameters
becomes a good solution for TDDFT to describe CT excitations, and the accuracy can
reach to 0.1 eV in the best case.\cite{stein2009reliable,tawada2004long,lange2008charge,minami2011nonempirically}
Other methods developed from DFT framework, including constrained DFT (CDFT)\cite{wu2005direct},
$\Delta$ self-consistent field approach ($\Delta$SCF)
\cite{ziegler1977calculation,jones1989density,gilbert2008self,barca2018simple},
and the perturbative $\Delta$SCF \cite{baruah2012charge,baruah2009dft,amerikheirabadi2018dft}
have been reported to provide accurate CT excitations energy as well,
and the computational cost can be maintained at DFT calculation level.

Similarly to TDDFT, particle-particle random phase approximation (pp-RPA) \cite{van2013exchange,van2014exchange,yang2013double}
can be viewed as a linear response theory for the time-dependent perturbation of
a pairing field on the ground states described with DFT. \cite{peng2014linear}
It has been shown to produce good CT results, as well as excellent description of
the asymptotically correct 1/R trend. \cite{yang2017charge}

Building beyond the framework of DFT, there are other methods successfully
developed to describe CT excitations. The many-body perturbation theory
at GW level and Bethe-Salpeter approach (BSE) \cite{leng2016gw,blase2011charge}
can yield results of similar accuracy as TDDFT from the best range-separated functional.
However, in terms of computational cost, these methods in general have computational
scaling higher or similar to TDDFT.

We would like to explore if it is possible to describe CT excitation
well directly from the ground state DFT calculation. 
    The (G)KS HOMO and LUMO  energies from density 
    functional approximation (DFA) with minimum 
    delocalization error are good approximations to $-I$ 
    and $-A$, based on the rigorously proven results on 
    connection to the chemical potentials. 
    \cite{cohenFractionalChargePerspective2008a} 
    We observed from 
    our recent work \cite{yuncaispectrum2018} that the remaining (G)KS 
    orbital energies from the localized orbital scaling correction (LOSC),
    a DFA with minimal delocalization error, \cite{li2017localized}
    approximate the corresponding quasiparticle energies well.
    Thus the entire set of (G)KS orbital energies from a DFA with
    minimal delocalization error can be used to obtain the photoemission spectrum
    with good agreement to the experimental reference, suggesting these orbital
    energies are good approximations to the quasiparticle energies.
    Based on this observation, we have demonstrated that the 
    excitation energy of an $N$-electron system can be approximated 
    by the orbital energy difference from an $(N-1)$-electron system. 
Bartlett\cite{haiduke2018communication}
and our group \cite{yuncaispectrum2018,yuncaispectrum2018arxiv}
demonstrated this is an efficient
way to calculate the excitation energy
by orbital energies from (G)KS-DFT. Results from our previous work
have showed good performance for the low-lying and Rydberg excitation energies
within this methodology. \cite{yuncaispectrum2018}
Therefore, in this paper, we further explore this approach for
application to CT excitation problems.

\section{Method and calculations}
Here, we briefly review the calculation method. 
Note here, we refer orbital energies to the ones either from KS calculations
for approximated exchange-correlation energy $E_{\rm{xc}}$ as an explicit
and continuous functional of electron density or from GKS calculations for
approximate $E_{\rm{xc}}$ as an explicit and continuous functional of the
non-interesting one-particle density matrix.
As we demonstrated in our recent work, orbital
energies $\varepsilon_m(N)/\varepsilon_n(N)$ from a DFA with minimal delocalization error are good approximation to the
quasiparticle/quasihole energies $\bm{\omega}^{+/-}\left(N\right)$ \cite{yuncaispectrum2018}; namely,
\begin{equation}
    \varepsilon_{m}(N) \approx\omega_{m}^{+}(N)=E_{m}(N+1)-E_{0}(N) \label{eq:ea}
\end{equation}
and
\begin{equation}
    \varepsilon_{n}(N) \approx\omega_{n}^{-}(N)=E_{0}(N)-E_{n}(N-1).\label{eq:ip}
\end{equation}
In Eq. \ref{eq:ea}, the virtual orbital energy, $\varepsilon_{m}(N)$,
is connected to a
one-electron addition process and approximates the electron affinity 
of the $N$-electron system. In Eq. \ref{eq:ip}, the occupied orbital
energy, $\varepsilon_{n}(N)$, is connected to a one-electron removal process 
and approximates the ionization potential of the $N$-electron system.
Utilizing these properties, the $m$th excitation energy $\Delta E_m\left(N\right)$,
defined as the energy difference between the $m$th excited states and ground state of
$N$-electron system, can be calculated from a ground-state $(N-1)$-electron system via
its virtual orbital energy difference, namely,
\begin{align}
    \label{eq:e_n-1}
    \Delta E_{m}(N) & =E_{m}(N)-E_{0}(N)\nonumber \\
                    & =[E_{m}(N)-E_{0}(N-1)]-[E_{0}(N)-E_{0}(N-1)]\nonumber \\
                    & =\omega_{m}^{+}(N-1)-\omega_{min}^{+}(N-1)\nonumber \\
                    & \approx\varepsilon_{m}(N-1)-\varepsilon_{LUMO}(N-1).
\end{align}
An alternative way to $(N-1)$-electron system calculation, one can switch to $(N+1)$-electron
system calculation and use occupied orbital energies instead.
However, the anionic $(N+1)$-electron
system normally are difficult to calculate and yields 
unreliable results \cite{yuncaispectrum2018}. Therefore we only focus on
the $(N-1)$-electron system calculation in this paper. In details, starting with a doublet
ground state $(N-1)$-electron system (assuming one more alpha electron than beta electron),
adding one electron to alpha orbital yields a triplet state, $E_{\uparrow\uparrow}(N)$,
which can be directly used for triplet excitation, and adding one electron
to beta orbital gives a spin-mixed state, $E_{\uparrow\downarrow}(N)$. In order to obtain singlet
excited state, commonly used spin purification process
\cite{ziegler1977calculation} are applied, i.e.
\begin{equation}
    \label{eq:spin_pure_E} E^{{\rm singlet}}(N) = 2E_{\uparrow\downarrow}(N) - E_{\uparrow\uparrow}(N),
\end{equation}
which leads singlet excitation energy to be expressed as
\begin{equation}
    \Delta E_{m}^{{\rm singlet}}\left(N\right)\approx
        [2\varepsilon_{m}^{\beta}\left(N-1\right)-\varepsilon_{m}^{\alpha}\left(N-1\right)] -
        \varepsilon_{{\rm LUMO}}^{\beta}\left(N-1\right).\label{eq:e_singlet}
\end{equation}

To test the performance of this methodology for CT excitations, we chose the standard
CT set from Stein $et$ $al$ \cite{stein2009reliable},
in which each system is composed of one strong electron withdrawing
molecule, tetracyanoethylene (TCNE), and one electron donating molecule ranging from benzene to
derivatives of anthracene, and their first excitation possesses clear CT characteristics from HOMO
to LUMO. In this work, we mainly focused on the calculation of the singlet excitation energy
and the description of $1/R$ trend for the test set via Eq. \ref{eq:e_singlet}.
We tested the performance of several conventional DFAs,
including LDA, \cite{slater1974quantum, vosko1980accurate}
BLYP\cite{becke1988density}, B3LYP \cite{becke1993becke,lee1988development},
and CAMB3LYP \cite{yanai2004new}, which are in a decreasing order of
delocalization error.
In addition, we tested the performance of these DFAs with
localized orbital scaling correction (LOSC) \cite{li2017localized}.
This correction method was recently developed in our group and
has been demonstrated to be capable of largely and systematically eliminating
delocalization error for conventional DFAs. Through this work, the LOSC calculation
was applied to the results from DFAs in a post-SCF approach. In particular,
the SCF calculation from conventional DFAs was performed first
to get the canonical orbitals and orbital energies, then 
a restrained Boys localization was applied to get a set of
special localized orbitals called orbitalets,
and only one-step correction, based on the orbitalets,
was added to the orbital energies at the end.
The correction process in LOSC can also be performed with
an SCF manner, however, applying post-SCF calculation in LOSC is more efficient
and the results have been demonstrated to be only slightly different from the ones
obtained by SCF process\cite{li2017localized}. Therefore, we only applied LOSC with the post-SCF
calculations in this work.
For all the DFT calculation in this work, we used our in-house developed package, QM4D \cite{qm4d}.
An unrestricted calculation was applied to all the test molecules. The basis set
used for DFT calculation is cc-pVTZ, and an auxiliary basis set,
used for density fitting in LOSC calculation\cite{li2017localized}, is aug-cc-pVTZ.

\section{Results and discussion}
To evaluate the performance of our methodology (Eq. \ref{eq:e_singlet}), we first test the
dependence of results on basis set. Table \ref{tab:converge} shows the singlet HOMO-LUMO
excitation energy of Xylene-TCNE complex from several common basis sets, and the results
were calculated from BLYP and LOSC-BLYP functional. According to this table, it suggests that
excitation energy from conventional DFA does not have much dependence on basis set choice.
For BLYP, the maximum excitation energy difference is only 0.1 eV. For results from LOSC-DFA,
the excitation energy keeps dropping, and the maximum difference was 0.27 eV for LOSC-BLYP.
Considering the cost and accuracy, we finally selected cc-pVTZ as the basis set for the
remaining calculations.

\begin{table}[]
    \caption{Basis set dependence for the singlet HOMO-LUMO excitation
            energy (in eV) of Xylene-TCNE complex, calculating from
            DFT with BLYP and LOSC-BLYP functionals.}
    \label{tab:converge}
    \begin{ruledtabular}
    \begin{tabular}{lcc}
    Basis              & BLYP  & LOSC-BLYP\\
    \hline
    cc-pVTZ            & 2.65 & 2.43 \\
    cc-pVQZ            & 2.64 & 2.33 \\
    6-31G*             & 2.74 & 2.55 \\
    6-31G**            & 2.73 & 2.54 \\
    6-31++G**          & 2.66 & 2.38 \\
    6-311++G**         & 2.65 & 2.36 \\
    6-311++G(3df, 3pd) & 2.63 & 2.28
    \end{tabular}
    \end{ruledtabular}
\end{table}

\subsection{Excitation energies}
To show the results, we start with the performance of CT excitation energy calculation.
Table \ref{tab:result_DFT} shows the results from tested DFAs and LOSC-DFAs via Eq. \ref{eq:e_singlet}.
In addition, we summarize results from other methods, including TD-DFT,
pp-RPA, CC2, $GW$-$BSE$, and $\Delta$SCF in Table \ref{tab:result_liter}.
Gas phase experimental data were used as reference. Among all the test cases,
there are some molecules, of which only the liquid phase experimental
data were applicable. As suggested in Ref. \citenum{stein2009reliable},
a universal correction (0.32 eV) was added to liquid phase data for the compensation of solvation energy,
hence, a complete set of gas phase reference was able to obtained.
According to Table \ref{tab:result_DFT}, we observed that the results from conventional DFAs show
mean absolute error (MAE) over 0.5 eV. Meanwhile, we noticed that MAE
generally shows a decreasing trend from 0.76 eV (LDA) to 0.56 eV (CAMB3LYP). This decreasing order
matches with the decreasing amount of delocalization error
in the approximate functionals from LDA to CAMB3LYP,
demonstrating the role of delocalization error in CT excitations within
this method. This idea was further supported by the results from LOSC-DFAs. As applying
LOSC can largely eliminate the delocalization error, the MAE of results from LOSC-DFAs
drops down as expected, especially for LOSC-LDA and LOSC-BLYP.
In particularly, the MAE from LOSC-LDA, the best case, can reach to 0.41 eV.
Comparing the results from our method with other calculation approaches shown in Table
\ref{tab:result_liter}, we noticed that our method cannot yield results with similar
accuracy to TDDFT with system-dependent tuned functional,
however, our method still performs better than TDDFT with conventional
functionals. Especially compared with TD-B3LYP (MAE of 1.06 eV),
the MAE from our method LOSC-B3LYP drops to 0.67 eV. We further
noticed that the performance of our method with LOSC-DFAs was similar
to $\Delta$SCF method. In the case of $\Delta$-BLYP and LOSC-BLYP from our method,
the MAE and mean signed error (MSE) are very similar.
Compared our results with $GW$-$BSE$ results, we observed that our best results are slightly
better than $G_0W_0$-$BSE$, but still worse than $GW$-$BSE$ with partial self-consistent calculations.
At the end, we noticed that most of the results from Table \ref{tab:result_DFT} and \ref{tab:result_liter}
show strong bias of underestimation to the experimental reference. This could be related to
the systematic error caused by the universal correction from the liquid phase experimental data.
Concerning this issue, we performed CC2 calculation, which is a high-level wavefunction method and
computationally affordable for this test set. Results from CC2 also shows 0.37 eV underestimation
to the experimental reference. Comparing our results directly with CC2, we found that our best results,
0.41 eV from LOSC-LDA, are similar.

\begin{turnpage}
\begin{table}[]
    \caption{Charge transfer excitation energies (in eV) obtained from DFT calculation with
             DFAs, LOSC-DFAs, comparing with experimental reference.}
    \label{tab:result_DFT}

    \begin{ruledtabular}
    \begin{tabular}{@{}lccccccccc@{}}
    System (TCNE)& Exp \footnotemark[1]&  LOSC-LDA & LOSC-BLYP & LOSC-B3LYP & LOSC-CAMB3LYP & LDA & BLYP & B3LYP & CAMB3LYP \\
    \hline
    Benezene     & 3.59 & NA    & NA    & 3.06  & 3.06  & NA    & NA    & 3.47  & 3.30  \\
    Toluene      & 3.36 & 2.79  & NA    & 2.64  & 2.65  & 2.79  & NA    & 2.96  & 2.85  \\
    Xylene       & 3.15 & 2.39  & 2.43  & 2.35  & 2.32  & 2.63  & 2.65  & 2.61  & 2.50  \\
    Naphtalene   & 2.60 & 2.20  & 2.08  & 2.02  & 2.24  & 2.12  & 2.06  & 2.05  & 2.14  \\
    \\
    Anthracene \footnotemark[2]\\
    None         & 2.05$^*$ & 1.74  & 1.78  & 1.44  & 1.64  & 1.27  & 1.21  & 1.30  & 1.50  \\
    cyano        & 2.33$^*$ & 2.10  & 1.57  & 1.48  & 1.81  & 1.36  & 1.32  & 1.43  & 1.66  \\
    chloro       & 2.06$^*$ & 1.65  & 1.63  & 1.37  & 1.58  & 1.19  & 1.14  & 1.24  & 1.44  \\
    carbomethoxy & 2.16$^*$ & 1.77  & 1.79  & 1.53  & 1.62  & 1.19  & 1.14  & 1.25  & 1.47  \\
    methyl       & 1.87$^*$ & 1.52  & 1.46  & 1.26  & 1.41  & 1.10  & 1.05  & 1.13  & 1.30  \\
    dimethyl     & 1.76$^*$ & 1.57  & 1.54  & 1.33  & 1.46  & 1.15  & 1.11  & 1.22  & 1.38  \\
    formyl       & 2.22$^*$ & 1.91  & 1.71  & 1.43  & 1.71  & 1.35  & 1.29  & 1.39  & 1.60  \\
    formychloro  & 2.28$^*$ & 1.74  & 1.67  & 1.40  & 1.70  & 1.34  & 1.28  & 1.40  & 1.62  \\
    \hline
    MAE          &      & 0.41  & 0.48  & 0.68  & 0.52  & 0.76  & 0.82  & 0.67  & 0.56  \\
    MSE          &      & -0.41 & -0.48 & -0.68 & -0.52 & -0.76 & -0.82 & -0.67 & -0.56
    \end{tabular}
    \end{ruledtabular}

    \footnotemark[1]{Experimental data refer to gas phase. Data was taken from Ref. \citenum{masnovi1984electron,hanazaki1972vapor}.
        If an asterisk was labeled to the experimental data, it refers to that the data was
        corrected from the liquid phase experimental data by 0.32 eV
        as suggested by Ref. \citenum{stein2009reliable}.}
    \footnotemark[2]{Systems shown below are substitutes of anthracene.}
\end{table}
\end{turnpage}

\begin{turnpage}
\begin{table}[]
    \caption{Charge transfer excitation energies (in eV) from TDDFT,
             pp-RPA, ${GW}$-${BSE}$, CC2 and $\Delta$SCF, comparing with experimental reference.}
    \label{tab:result_liter}

    \begin{ruledtabular}
    \begin{tabular}{@{}lcccccccc@{}}
    System (TCNE)& Exp\footnotemark[1] & TD-tuned-BNL \footnotemark[2]& TD-B3LYP\footnotemark[2]& pp-RPA \footnotemark[3]   & CC2 \footnotemark[4]
        & $G_0W_0$-$BSE$\footnotemark[5]& $GW_{psc}$-$BSE$\footnotemark[5] & BLYP-MOM\footnotemark[6]\\
    \hline
    Benezene     & 3.59   & 3.80    & 2.10     & NA     & 3.40  & 3.03     & 3.58   &NA \\
    Toluene      & 3.36   & 3.40    & 1.80     & NA     & 3.03  & 2.52     & 3.27   &2.67 \\
    Xylene       & 3.15   & 3.00    & 1.50     & NA     & 2.75  & 2.23     & 2.89   &3.23 \\
    Naphtalene   & 2.60   & 2.70    & 0.90     & 2.27   & 2.29  & 1.96     & 2.55   &3.22 \\
    \\
    Anthracene \footnotemark[7]\\
    None         & 2.05$^*$   & 2.14    & 1.32     & 1.62   & 1.65  & 1.66     & 2.17   &NA  \\
    cyano        & 2.33$^*$   & 2.35    & 0.82     & 1.74   & 1.87  & 1.89     & 2.32   &1.85  \\
    chloro       & 2.06$^*$   & 2.14    & 1.32     & 1.58   & 1.65  & NA       & NA     &1.57  \\
    carbomethoxy & 2.16$^*$   & 2.16    & 1.22     & 1.58   & 1.69  & 1.64     & 2.05   &1.61  \\
    methyl       & 1.87$^*$   & 2.03    & 1.42     & 1.39   & 1.49  & 1.21     & 1.99   &1.46  \\
    dimethyl     & 1.76$^*$   & 2.09    & 1.72     & 1.41   & 1.53  & 2.00     & 2.21   &1.43  \\
    formyl       & 2.22$^*$   & 2.27    & 1.32     & 1.78   & 1.81  & 1.89     & 2.32   &1.76  \\
    formychloro  & 2.28$^*$   & 2.28    & 1.22     & 1.79   & 1.85  & NA       & NA     &1.78  \\
    \hline
    MAE          &        & 0.10    & 1.06     & 0.46   & 0.37  & 0.55     & 0.13   &0.46  \\
    MSE          &        & 0.08    & -1.06    & -0.46  & -0.37 & -0.51    & 0.03   &-0.32
    \end{tabular}
    \end{ruledtabular}

    \footnotemark[1]{Experimental data refer to gas phase. Data was taken from Ref. \citenum{masnovi1984electron,hanazaki1972vapor}.
                     If an asterisk was labeled to the experimental data, it refers to that the data was
                     corrected from the liquid phase experimental data by 0.32 eV
                     as suggested by Ref. \citenum{stein2009reliable}.}
    \footnotemark[2]{Data taken from Ref. \citenum{stein2009reliable}.}
    \footnotemark[3]{Results from pp-RPA with CAMB3LYP reference. Data were taken from Ref \citenum{yang2017charge}.}
    \footnotemark[4]{CC2 calculation was performed by package Turbomole \cite{TURBOMOLE}. Aug-cc-pVTZ was used as basis set,
                     which was tested to reach the basis set convergence.}
    \footnotemark[5]{Data were taken from Ref \citenum{blase2011charge}. $GW_{psc}$ stands for partial self-consistent $GW$.}
    \footnotemark[6]{$\Delta$SCF-BLYP with maximum overlap method \cite{gilbert2008self}.}
    \footnotemark[7]{Systems shown below are substitutes of anthracene.}
\end{table}
\end{turnpage}

\subsection{$1/R$ dependence analysis}

\begin{figure}[htbp]
    \centering
    \includegraphics[width=\textwidth]{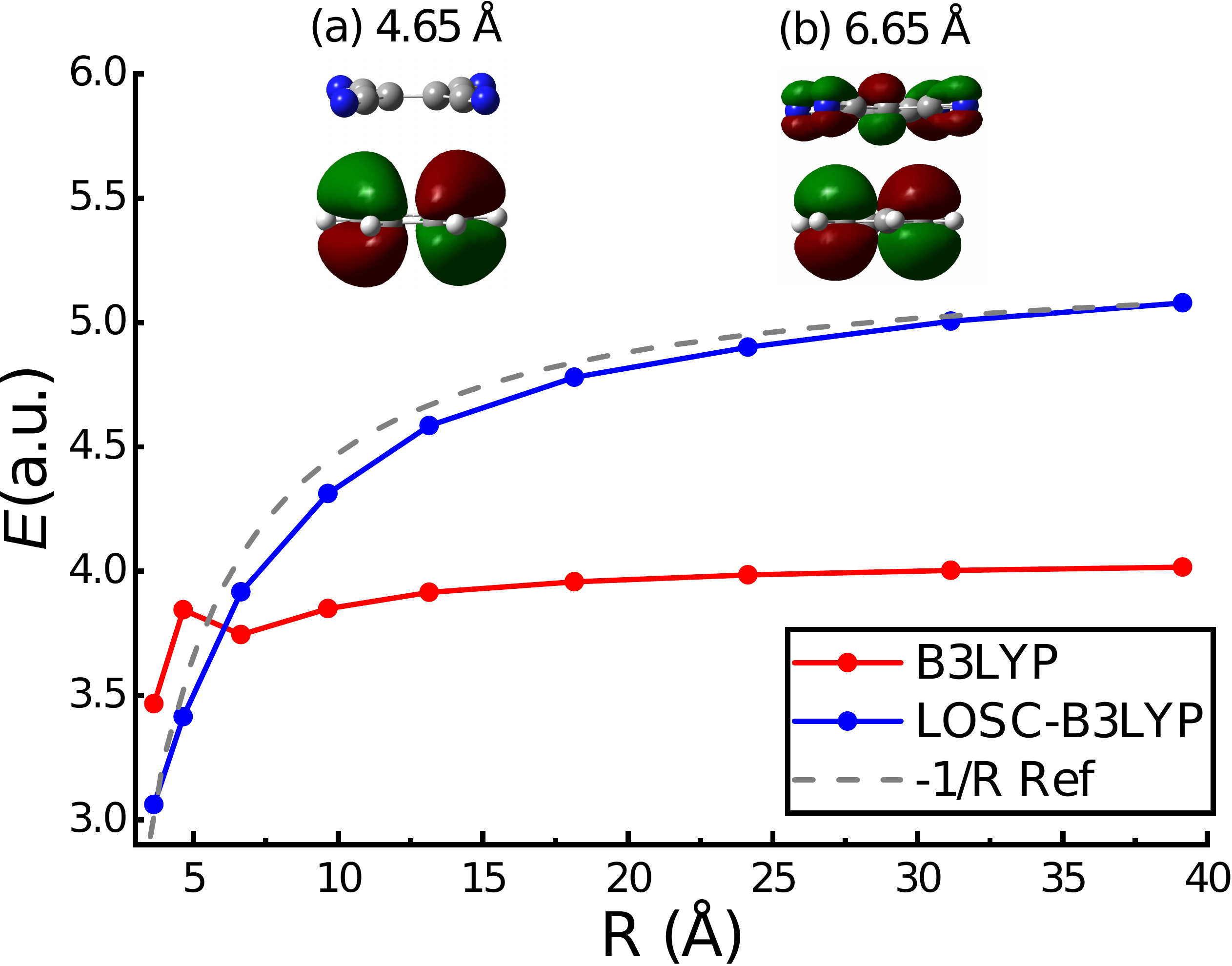}
    \caption{Variation of the benzene-TCNE complex first singlet charge transfer excitation
             energies (in eV) with intermolecular distance $R$. Subfigure (a) and (b)
             show the electron distribution of LUMO orbital of the $(N-1)$-electron benzene-TCNE complex system with
             increasing intermolecular distance at 4.65 and 6.65 \text{\normalfont\AA}, respectively.}
    \label{fig:1/R}
\end{figure}

Followed by showing the CT excitation energies, we now verify
the description of $1/R$ dependence from this method. The excitation
energy curve of benzene-TCNE complex with respect to the variation of intermolecular distance
was plotted in Fig. \ref{fig:1/R}. By observing Fig. \ref{fig:1/R},
we noticed that applying B3LYP fails to describe
the $1/R$ behavior. Especially at long distance, the curve from B3LYP
in Fig. \ref{fig:1/R} shows a critical drop starting around
5 \text{\normalfont{\AA}}.
The reason for the poor performance of B3LYP can be mainly attributed to
the delocalized orbital density at long distance.
To clarify this point, figure (a) and (b) of Fig. \ref{fig:1/R}
show the LUMO electron distribution of $(N-1)$-electron system
with different separation distances at 4.65 and 6.65 \text{\normalfont{\AA}}.
According to these figures, 
the LUMO of the $(N-1)$-electron system starts to delocalize from
the donor (benzene) to the acceptor (TCNE), when the separation
increases to large distance.
Since the LUMO of the $(N-1)$-electron system is
used to retrieve the ground state of $N$-electron system,
such orbital delocalization to acceptor molecule leads only partial,
rather than one-net, electron transition within this method,
and brings the calculated CT excitation energies lower than expected.
When LOSC is applied to B3LYP to reduce the delocalization error,
it shows that $1/R$ behavior can be correctly captured.

\begin{figure}[htbp]
    \centering
    \includegraphics[width=\textwidth]{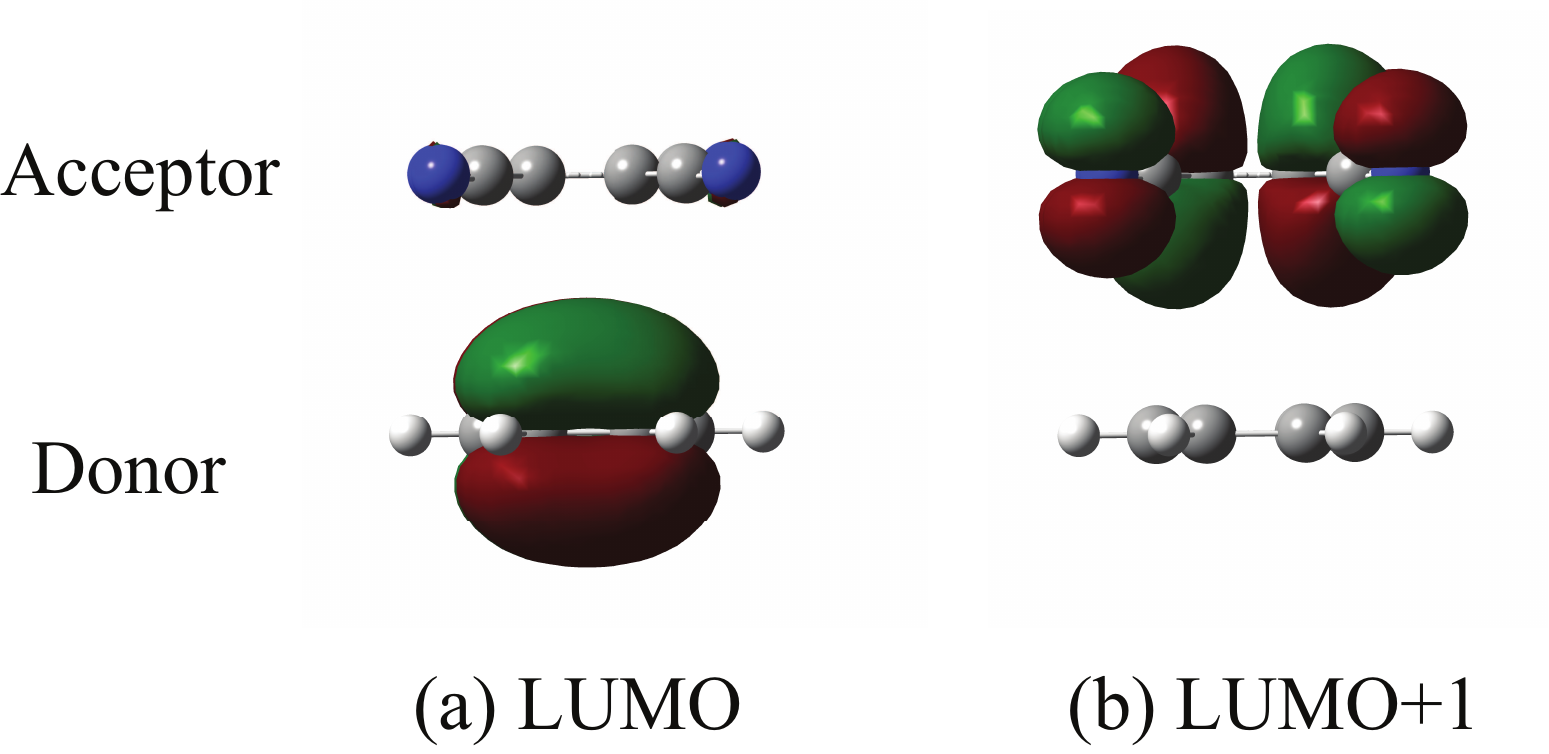}
    \caption{Electron distribution of (a) LUMO and (b) LUMO+1 orbital of the $(N-1)$-electron benzene-TCNE complex system
             from B3LYP calculation at short intermolecular distance, 3.65 \text{\normalfont\AA}.}
    \label{fig:e_dst}
\end{figure}

To understand the reason that the $1/R$ trend can only be captured
from DFA with minimal delocalization error within this methodology,
we start analysis on a model donor-acceptor system.
First, the CT excitation of the donor-acceptor complex is concerning the transition of
one electron from the donor molecule to the acceptor molecule.
When we start the calculation from the ground state $(N-1)$-electron
system, it would be reasonable
to assume one electron was completely removed from the donor molecule only.
This assumption of $(N-1)$-electron system will make the donor molecular become one positively charged, and
leave the acceptor molecule to stay neutral. Under this condition, the virtual orbital
from such $(N-1)$-electron system, in particular, the LUMO orbital should be
localized on the donor molecular, since adding one electron to LUMO gives
the ground state $N$-electron system. In contrast, the orbitals above LUMO
from $(N-1)$-electron system should be localized on acceptor molecular,
since adding one electron to these orbitals gives the CT excited state of $N$-electron system.
This consideration of the localization for virtual orbitals from $(N-1)$-electron
system was confirmed to be reasonable by plotting the electron distribution of orbitals from one
test case, benzene-TCNE complex. As shown in Fig. \ref{fig:e_dst}, LUMO of
the complex at short separation distance (3.65 \text{\normalfont\AA}) is
localized on benzene molecule (donor), while LUMO+1 is localized on the TCNE molecule (acceptor).

Based on the above assumption that the concerning orbitals from $(N-1)$-electron system are localized
in corresponding subspace,
the positive charge on the donor molecule acts as an external +1 point charge
field to the neutral acceptor molecule. Therefore,
the orbitals above LUMO, which are localized on the acceptor molecule,
should be lowered by $1/R$. In contrast, the LUMO, which is localized on
the donor molecule, should have negligible effect from the neutral acceptor molecule, if
we ignore the higher-order polarization effect.
According to this analysis, the CT excitation energy
from this method, (Eq. \ref{eq:e_singlet}), should have the dependence of
\begin{equation}
    \label{eq:1/R_std}  \Delta E_m^{\rm singlet}  \propto [-2/R - (-1/R)] - 0 = -1/R,
\end{equation}
if the orbital localization condition holds.

As conventional DFAs suffer from delocalization error seriously and
the orbitals start to delocalize at long distance,
$1/R$ trend as described in Eq. \ref{eq:1/R_std} will not be expected
from conventional DFAs. In the case of LOSC calculation, if the correction was performed with an SCF manner,
the orbitals will be expected to be localized from conventional DFAs (see Ref. \citenum{li2017localized}),
hence, the $1/R$ trend would be expected in principle.
If the LOSC calculation was only performed with a post-SCF manner, which is the case in this work,
we still observed the qualitatively $1/R$ trend. Although
the post-SCF LOSC calculation does not update the orbital density information
from parent DFAs, it may still be delocalized at long distance. It turns out that
orbital energies are corrected accurately by LOSC to the quasiparticle energies. Therefore,
the CT excitation energy, Eq. \ref{eq:CT}, is well approximated by Eq. \ref{eq:e_singlet}
numerically, and the $1/R$ dependence is observed.

\section{conclusion}
In conclusion, we demonstrated a simple and efficient way to
calculate the CT excitation energy of an $N$-electron system from
the Kohn-Sham orbital energies of a ground state $(N-1)$-electron system.
With numerical results, we show that our method performs better in CT excitation energy calculation
to the TDDFT with conventional functionals. To obtain reliable results within this method,
applying density functional approximation with minimal delocalization error, such LOSC-DFAs, are important. The computational
cost from this method is only at the level of a DFT ground state
calculation. However, the accuracy can reach to a degree similar
to the $\Delta$SCF, $G_0W_0$-$BSE$, pp-RPA and even close to
high-level wavefunction method, like CC2.
In addition to the CT excitation energy calculations, we verified
that this method is capable of capturing the $1/R$ trend of CT excitation, in which applying LOSC-DFAs
is essential. Analysis shows that the $1/R$ trend is always expected from this method, as long as the orbital localization
condition holds. 
Considering the good results and inexpensive
computational cost from the ground state DFT calculations,
we believe this method provides an efficient path to describe
CT excitation problems.

\section{Acknowledgement}
Support from the National Institutes of Health (Grant No. R01 GM061870-13)
(WY), the Center for Computational Design of Functional Layered Materials
(Award DE-SC0012575), an Energy Frontier Research Center funded by
the US Department of Energy, Office of Science, Basic Energy Sciences (YM).
\bibliography{reference,CT}
\end{document}